\documentstyle[epsfig]{camera}
\begin{document}
\setcounter{topnumber}{5}
\def\topfraction{1.0}
\setcounter{bottomnumber}{5}
\def\bottomfraction{1.0}
\setcounter{totalnumber}{5}
\def\textfraction{0.0}
\def\floatpagefraction{1.0}
\pagenumbering{arabic} \setcounter{page}{1}

\def \met {{\,/\!\!\!\!E_{T}}}
\def \notbjet {{\check{j}}}
\def \bjet {b}
\def \deltaR {\Delta {\cal R}}
\newcommand{\et}{\rm E_T}
\def \code {\texttt}
\def \GeV {{\rm GeV}}
\def \metvec {{\rm\,/\!\!\!\!\vec{E}_{T}}}
\def \begineq {\begin{equation}}
\def \endeq {\end{equation}}
\def \scriptP {\mbox{${\cal P}$}}
\def \gothicP {\tilde{\scriptP}}
\def \ltapprox {\,\raisebox{-0.6ex}{$\stackrel{<}{\sim}$}\,}
\def \gtapprox {\,\raisebox{-0.6ex}{$\stackrel{>}{\sim}$}\,}
\def \Sherlock {{Sleuth}}
\def \hse {{\small{hse}}}
\newcommand {\abs}[1]{\mid \! #1 \! \mid}
\renewcommand{\thefootnote}{\arabic{footnote}}

\hyphenation{straight-forward}
\newcommand{\Pt}{$\rm P_T$}
\newcommand{\Pte}{$\rm P_T^e$}
\newcommand{\Et}{$\rm E_T$}
\newcommand{\Etj}{$\rm E_T^{j1}$}
\newcommand{\Etjj}{$\rm E_T^{j2}$}
\newcommand{\Etjjj}{$\rm E_T^{j2}+E_T^{j3}$}
\newcommand{\ppbar}{p{\bar p}}
\newcommand{\ttbar}{$t{\bar t}$}
\newcommand{\dytt}{Drell-Yan $\rightarrow \tau^+\tau^-$}
\newcommand{\dyll}{Drell-Yan $\rightarrow \ell^+\ell^-$}
\newcommand{\dyee}{Drell-Yan $\rightarrow e^+e^-$}
\newcommand{\ztt}{$Z^0 \rightarrow \tau^+\tau^-$}
\newcommand{\zee}{$Z^0 \rightarrow e^+e^-$}
\newcommand{\zmm}{$Z^0 \rightarrow \mu^+\mu^-$}
\newcommand{\DO}{$D\O$}
\newcommand{\pbarp}{p{\bar p}}
\newcommand{\etal}{{\em et al.}}
\newcommand{\mett}{\mbox{${\rm \not\! E}_{\rm T}$}}
\newcommand{\mettcal}{\mbox{${\rm \not\! E}_{\rm T}^{\rm cal}$}}

 \newcommand{\EMMDA}{39}
 \newcommand{\EMMFA}{18.4$\pm$1.4}
 \newcommand{\EMMTT}{0.011$\pm$0.003}
 \newcommand{\EMMDY}{0.5$\pm$0.2}
 \newcommand{\EMMWW}{3.9$\pm$1.0}
 \newcommand{\EMMZT}{25.6$\pm$6.5}
 \newcommand{\EMMTOT}{48.5$\pm$7.6}
 \newcommand{\EMMJDA}{13}
 \newcommand{\EMMJFA}{8.7$\pm$1.0}
 \newcommand{\EMMJTT}{0.4$\pm$0.1}
 \newcommand{\EMMJDY}{0.1$\pm$0.03}
 \newcommand{\EMMJWW}{1.1$\pm$0.3}
 \newcommand{\EMMJZT}{3$\pm$0.8}
 \newcommand{\EMMJTOT}{13.2$\pm$1.5}
 \newcommand{\EMMJJDA}{5}
 \newcommand{\EMMJJFA}{2.7$\pm$0.6}
 \newcommand{\EMMJJTT}{1.8$\pm$0.5}
 \newcommand{\EMMJJDY}{0.012$\pm$0.006}
 \newcommand{\EMMJJWW}{0.18$\pm$0.05}
 \newcommand{\EMMJJZT}{0.5$\pm$0.2}
 \newcommand{\EMMJJTOT}{5.2$\pm$0.8}
 \newcommand{\EMMJJJDA}{1}
 \newcommand{\EMMJJJFA}{0.4$\pm$0.2}
 \newcommand{\EMMJJJTT}{0.7$\pm$0.2}
 \newcommand{\EMMJJJDY}{0.005$\pm$0.004}
 \newcommand{\EMMJJJWW}{0.032$\pm$0.009}
 \newcommand{\EMMJJJZT}{0.07$\pm$0.05}
 \newcommand{\EMMJJJTOT}{1.3$\pm$0.3}
 \newcommand{\EMMJJJJDA}{0}
 \newcommand{\EMMJJJJTT}{0.16$\pm$0.04}
 \newcommand{\EMMJJJJDY}{0.002$\pm$0.003}
 \newcommand{\EMMJJJJWW}{0.004$\pm$0.002}
 \newcommand{\EMMJJJJZT}{0.02$\pm$0.03}
 \newcommand{\EMMJJJJTOT}{0.2$\pm$0.2}
 \newcommand{\EMMJJJJJDA}{0}
 \newcommand{\EMMJJJJJFA}{0$\pm$0.2}
 \newcommand{\EMMJJJJJTT}{0.025$\pm$0.007}
 \newcommand{\EMMJJJJJDY}{0$\pm$0.003}
 \newcommand{\EMMJJJJJWW}{0$\pm$0.0006}
 \newcommand{\EMMJJJJJZT}{0$\pm$0.03}
 \newcommand{\EMMJJJJJTOT}{0.025$\pm$0.2}
 \newcommand{\EMXDA}{58}
 \newcommand{\EMXFA}{30.2$\pm$1.8}
 \newcommand{\EMXTT}{3.1$\pm$0.5}
 \newcommand{\EMXDY}{0.7$\pm$0.1}
 \newcommand{\EMXWW}{5.2$\pm$0.8}
 \newcommand{\EMXZT}{29.2$\pm$4.5}
 \newcommand{\EMXTOT}{68.3$\pm$5.7}

\title{PROBING PHYSICS BEYOND THE SM AT TEVATRON}
%
\author{Carmine Pagliarone\footnote{pagliarone@fnal.gov}   \\
{\it (On behalf of CDF and  D$\not$O  Collaborations)}\\
INFN Sez. Pisa\\ {\it via Livornese 1291 - 56010 S.Piero a Grado (PI) - ITALY}}
\maketitle

\vspace{-2.5cm}
\begin{abstract}
Tevatron Experiments: CDF and  D$\not$O collected during
October 1992 and February 1996 (Run I) a data sample of roughly $120$
$pb^{-1}$ $ p \bar{p}$ collisions at a center of mass energy
$\sqrt{s}=$ $1.8$ TeV. A large variety of physical studies
have been performed using these data. Current
paper reviews last results obtained searching for physics
beyond the Standard Model. Direct Supersymmetry (SUSY) searches are not part of
this review.
\end{abstract}

\section{Introduction}
In this paper we review results obtained recently by CDF and D$\not$O
collaborations in the field of non-SUSY searches for new phenomena.
The analysis described below are based on Run I Tevatron data ($\sim$
$120$ $pb^{-1}$).\\
\noindent
Detailed description of Run I CDF and D$\not$O detectors
can be found in the following reference~\cite{CDFDO}.

\section{Search for Large Extra Dimensions}
The Standard Model (SM) has proved to be enormously successful in
providing a description of particle physics up to energy scales of
several hundred GeV as probed by current
experiments~\cite{SM-exp}. In the SM, however, one assumes that
effects of gravity can be neglected, because the scale where such
effects become large is the Planck Scale. The question of why the
$4-$dimensional Planck Scale, $G^{-1/2}_{Pl} \sim 10^{19}$ GeV, is
much larger than the electroweak (EWK) scale,
$G^{-1/2}_{{\mathcal{F}}} \sim 10^{2}$ GeV, is an outstanding
problem in contemporary physics.
Motivated in part by naturalness issues, numerous
scenarios have emerged recently, that address the hierarchy
problem within the context of the old idea that some part of the
physical world (i.e. the SM-world) is confined to a brane in a
higher dimensional space~\cite{kaluza-klein}.
As we don't experience more then $3$ spatial dimensions, we have
to assume that any possible Extra Spatial Dimension ($ESD$) is hidden i.e. compactified.

\noindent
The impact of virtual gravitons in hadron collider
experiments can be observed in processes such as: $q \bar{q}
\rightarrow G \rightarrow \gamma \gamma$ or $g g \rightarrow G
\rightarrow e^{+}e^{-}$ where the ADD model introduces production
mechanism that can increase the cross-section of diphoton and
di-electron production at high invariant mass over the SM. The
diphoton and di-electron cross-section considering the Large Extra
Dimension ($LED$) contributions take the form~\cite{HDD}:
\begin{equation}
\frac{d^{2}\sigma_{Tot}}{d cos\theta^{*}\,\,  dM} =
\frac{d^{2}\sigma_{SM}}{d cos\theta^{*}\,\,  dM} \,\,+\,\,
\frac{a(n)}{M^{4}_{\mathcal{F}}} \,\,F_{1}(cos \theta^{*}, M) \,\,+\,\,
\frac{b(n)}{M^{8}_{\mathcal{F}}} \,\,F_{2}(cos \theta^{*}, M)
\label{ccc}
\end{equation}
\begin{figure} [ht!] \centering
\hspace{-0.6cm}
\begin{minipage}{0.5\linewidth}
  \centering\epsfig{file=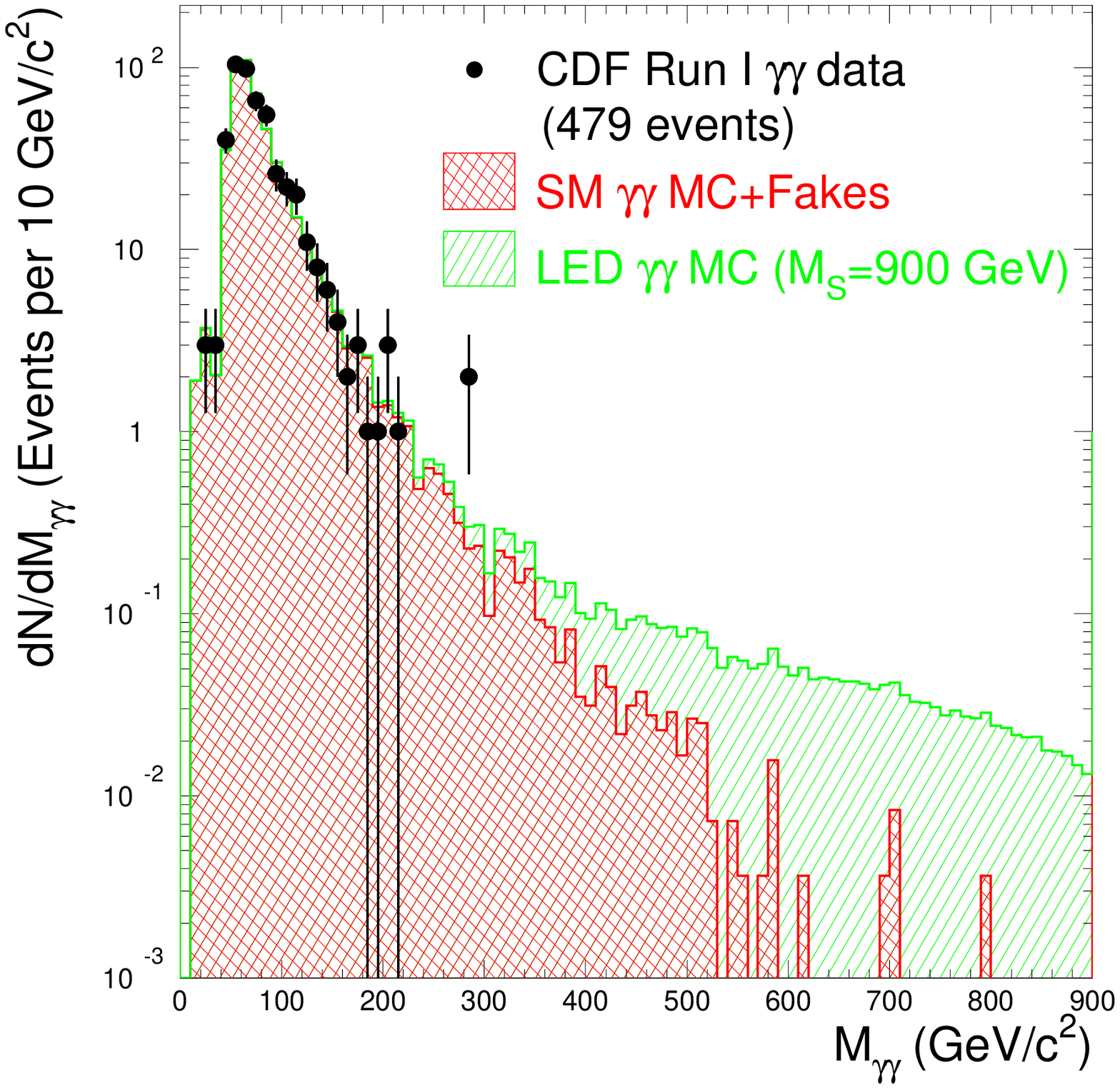,width=\linewidth}
\end{minipage}\hfill
\hspace{-0.5cm}
\begin{minipage}{0.47\linewidth}
  \centering\epsfig{file=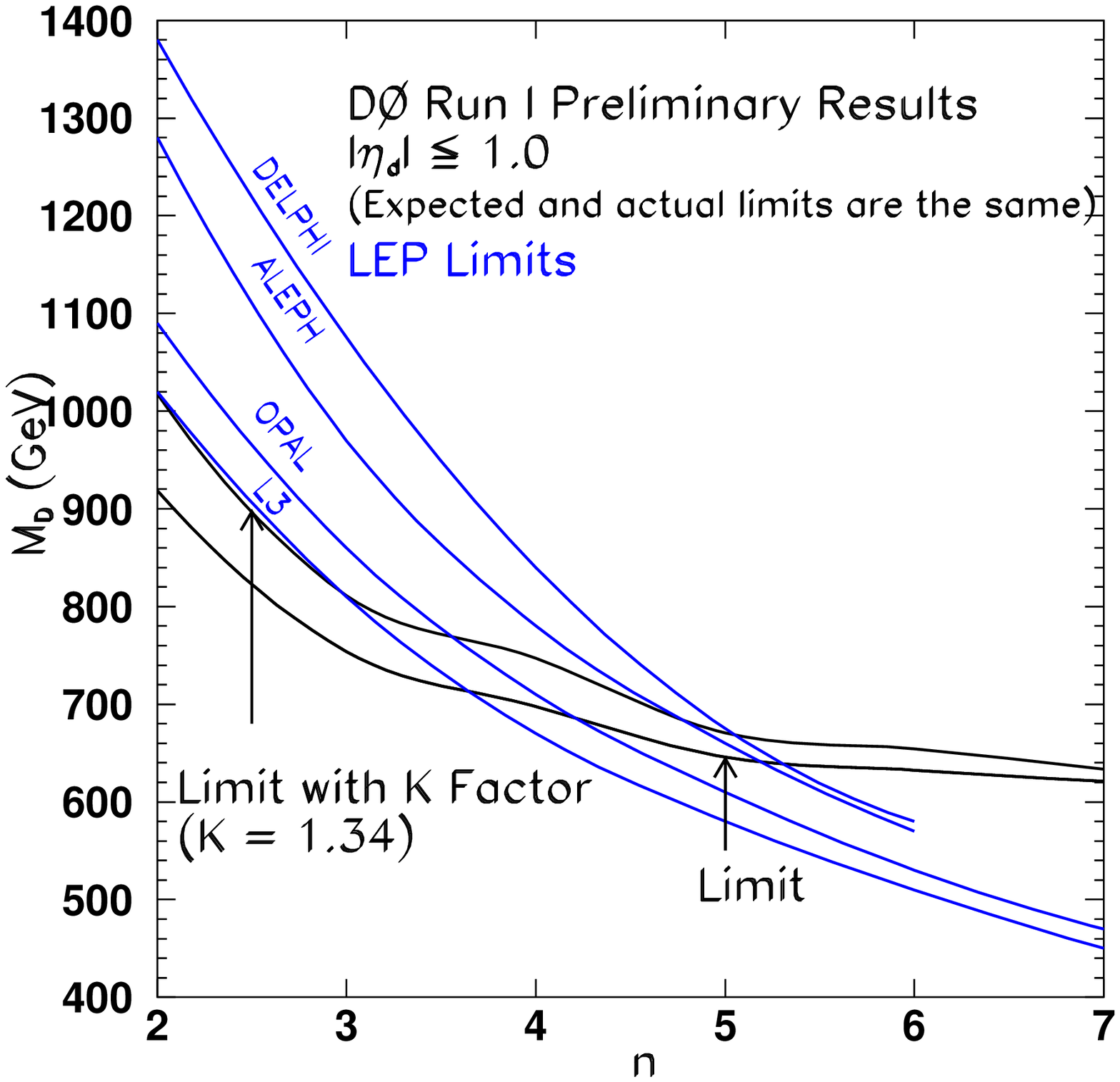,width=\linewidth}
\end{minipage}
\vspace{-.3cm}
  \caption{\it a) CDF Diphoton invariant mass distribution for data (dots),
SM $\gamma \gamma+$ fake (red area) and for the
LED contribution (green area); b)
95\% C.L. lower limits on $M_{D}$ in the D$\not$O
search for real graviton emission.}
\label{fig:CDFfig}
\end{figure}

\noindent
where $cos\theta^{*}$ is the scattering angle of $\gamma$ or $e$
in the center of mass frame of the incoming parton.
The first term in the expression~\ref{ccc} is the pure SM contribution
to the cross section; the second and the third part are
the interference term and the direct $G_{KK}$ contribution.
The characteristic signatures for contributions from virtual $G_{KK}$
correspond to the formation of massive systems abnormally beyond
the SM expectations. Figure~\ref{fig:CDFfig}.a shows a comparison of
the diphoton invariant mass for signal, background processes and
for the data. With no excess apparent beyond SM expectations
CDF proceeds to calculate a lower limit on the graviton
contribution to the di-electron, diphoton cross section. The limits are given
in table~\ref{tbl:comblimits}~\cite{SIMONA} and they are translated in the three canonical
notations: Hewett, GRW and HLZ~\cite{LEDNOT}.

\vspace{-.9cm}
\begin{table}[h!]
\label{tbl:comblimits}
\begin{center}
\vspace{.5cm}
\begin{tabular}{|c|c|cc|c|crccc|}
\hline \hline
&  &
\multicolumn{2}{c|}{\bf{Hewett}} & {\bf GRW} & \multicolumn{5}{c|}{\bf{HLZ}}\\ \hline
{\bf {Sample}} & ${\bf K}$ & $\lambda$=-1 & $\lambda$=+1 & & $n$=3 & $n$=4 & $n$=5 &
$n$=6 & $n$=7\\ \hline\hline
CC+CP $\gamma\gamma$ & 1.0 & 899 & 797 & 1006 & 1197 & 1006 & 909 & 846 &
800 \\$e^+e^-$ & 1.0 & 780 & 768 & 873 & 1038 & 873 & 789 & 734 & 694 \\
$e^+e^-$+$\gamma\gamma$ & 1.0 & 905 & 826 & 1013 & 1205 & 1013 & 916 & 852
& 806 \\
$e^+e^-$+$\gamma\gamma$ & 1.3 & 939 & 853 & 1051 & 1250 & 1051 & 950 & 884
& 836 \\
\hline \hline
\end{tabular}
\end{center}
\vspace{-.5cm}
\caption{CDF 95\% C.L. limit on the size of  ${ M_S}$ for $ee$, $\gamma \gamma$ and combined channels.
Combined results ($e^+e^-$+$\gamma\gamma$) have been determined for both $K=1.0$ and $K=1.3$.}
\end{table}
\noindent
Both CDF and D$\not$O Collaborations at the Fermilab
looked also for direct graviton emission. From
an experimental point of view the mono-jet plus missing$-\not\!\!\!E_{\rm T}$
signature is quite complex to study
because of the large instrumental background from jet
mismeasurement and the presence of cosmic rays background.
The D$\not$O 95\% C.L. limit on $M_{D}$ is given in Fig.~\ref{fig:LQcombined}.b.

\vspace{-.2cm}
\section{Leptoquarks}

\begin{figure} [t!] \centering
\begin{minipage}{0.5\linewidth}
  \centering\epsfig{file=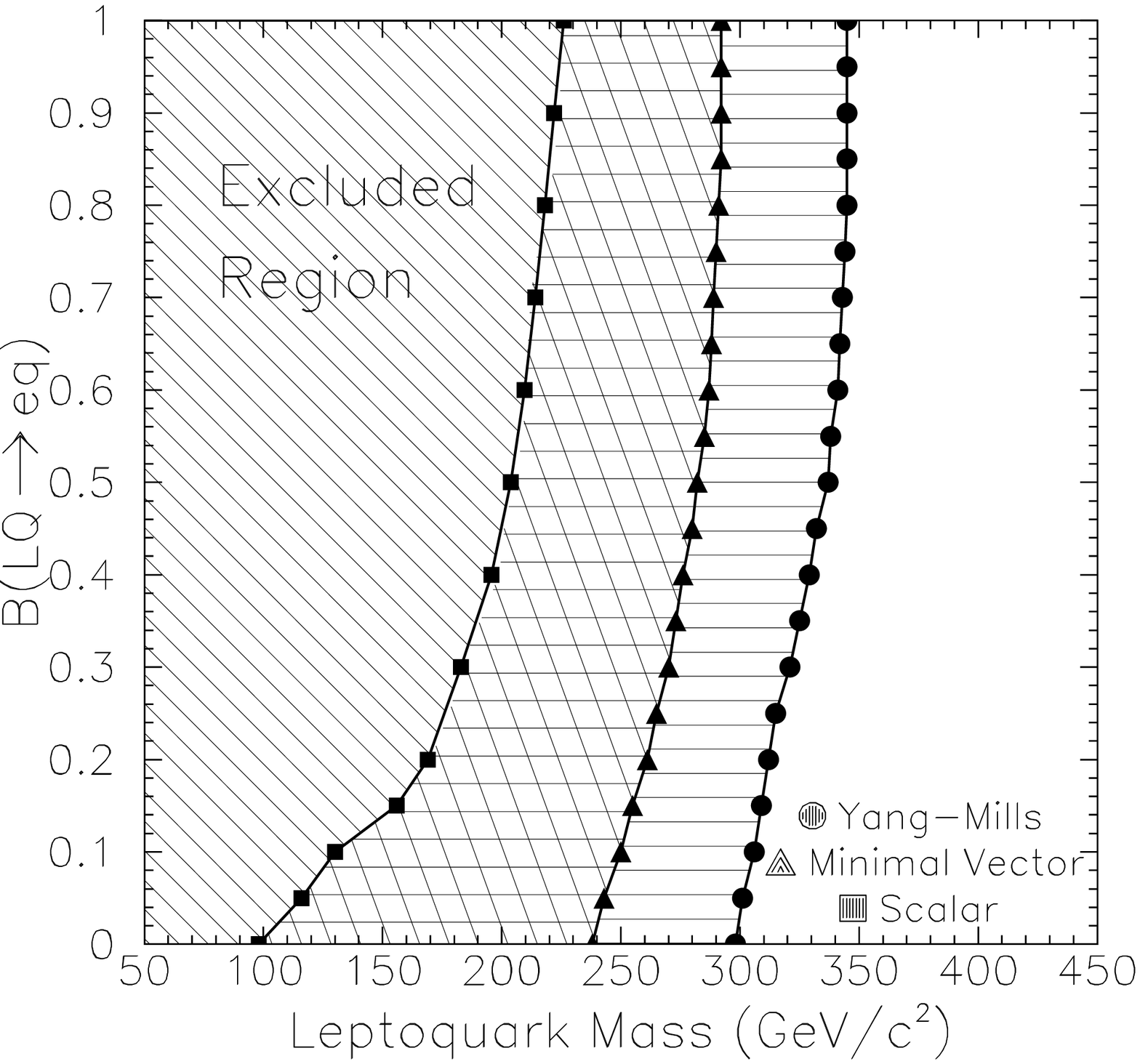,width=\linewidth}
\end{minipage}\hfill
\begin{minipage}{0.5\linewidth}
  \centering\epsfig{file=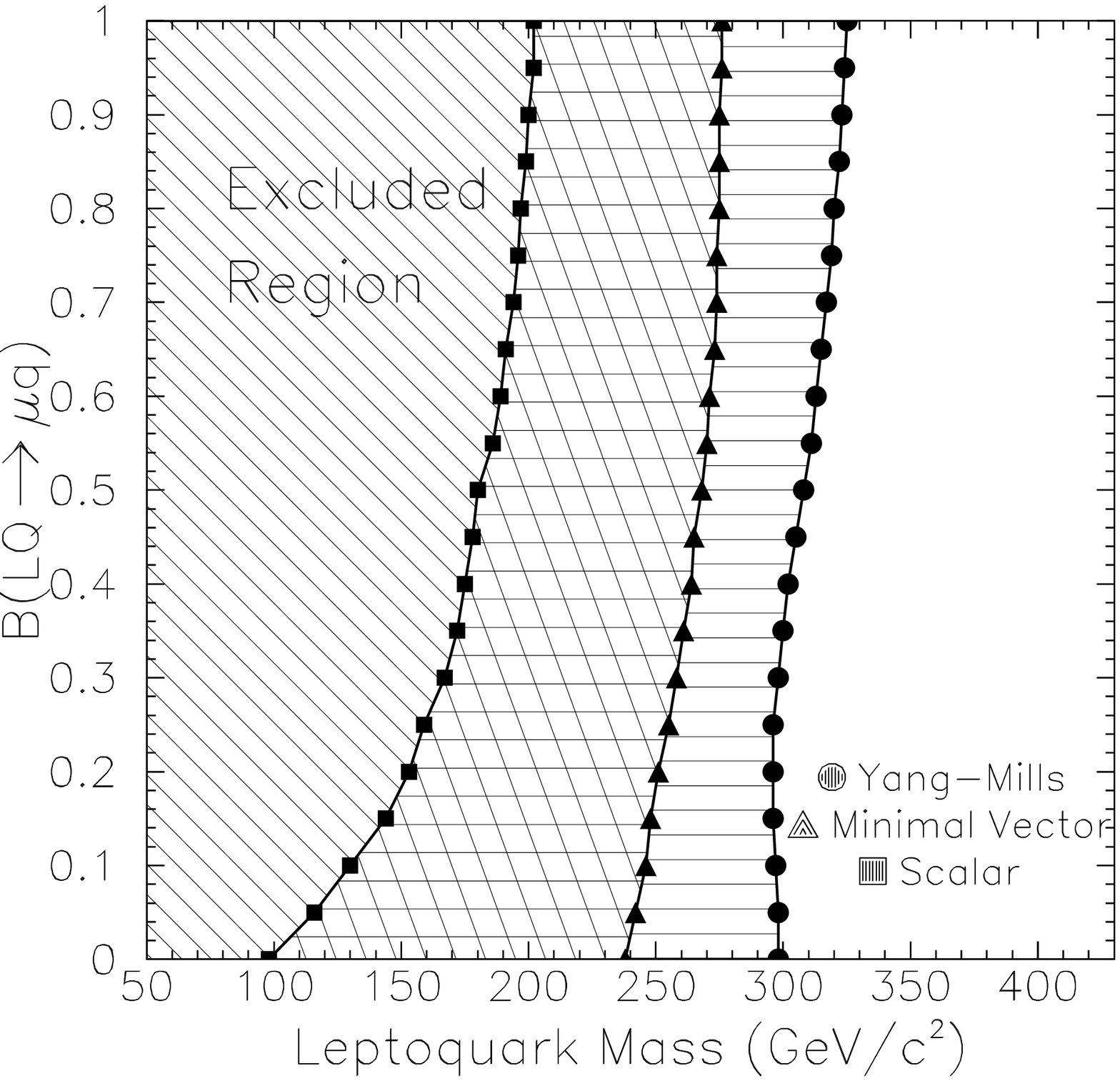,width=\linewidth}
\end{minipage}
\vspace{-.3cm}
  \caption{\it D$\not$O combined 95\% C.L. limit on the mass of scalar leptoquarks as a
function of BR(${\rm LQ} \rightarrow  l^{\pm}q$); a) for Scalar leptoquarks and
b) for Vector letpoquarks.}
  \label{fig:LQcombined}
\end{figure}

Leptoquarks (LQ) are predicted in many extensions of the SM such as:
Grand Unified Theories, Technicolor, etc.
At Tevatron they can be pair produced through strong
interactions: $p \bar{p} \rightarrow \bar{LQ}LQ + X$ and decay in one of
the following final states:
$\ell^{\pm}\ell^{\mp}q\bar{q}$ and $\ell^{\pm} \nu q \bar{q}$ and
$\nu \bar{\nu} q \bar{q}$.
Both Tevatron experiments searched in the past
for $LQ$ by looking at final states containing one or two leptons~\cite{CDFLQ,D0LQ}.
Here we report the latest D$\not$O analysis performed by
looking at the channel $\nu \bar{\nu} q \bar{q}$.
The main sources of background for this process are
SM multijet, $W+$Jets, $Z+$jets and $t \bar{t}$ processes.
Fig.~\ref{fig:LQcombined}.a and~\ref{fig:LQcombined}.b show
the 95\% C.L. limit obtained combining
the present analysis with the previous D$\not$O searches for both scalar and vector Leptoquarks.

\section{Model independent probes}

Until the number of compelling candidate theories of the {\it Nature}
was small, it was natural to try to rule out each theoretical
scenario, by finding observables that could truly help do differ
SM processes from what expected from the specific model under investigation.
Unfortunately, because of the complexity of the models, the
increasing number, as well as the large parameter space that very often
have to be considered for each of them, to follow a classical approach
may not be an economic way to investigate the nature.
In the past years the D0 collaboration explored different
approaches to the data analysis and come out with two useful
tools: SLEUTH and then QUAERO.\\
\noindent
SLEUTH is a quasi-model-independent search strategy for new
high $P_{T}$ physics. Given a data sample, its final state and
a set of variables to that final state (see table~\ref{tbl:variables}),
SLEUTH determines the most interesting region in those variables
and quantifies the degree of interest. The published results are
avaliable in the Ref.~\cite{SLEUTH}.\\
\noindent
QUAERO is a method that enables the automatic optimization of searches
for physics beyond the SM, providing a tool for making the
data avaliable to a larger public ({\it http://quaero.fnal.gov}).
QUAERO have been used in eleven separate searches such as
leptoquark production: $LQ\overline{LQ} \rightarrow ee2j$,
$W'$ and $Z'$ production: $W' \rightarrow WZ
\rightarrow e \met 2j$, $Z' \rightarrow t\overline{t} \rightarrow
e \met 4j$ and
SM higgs production: $h \rightarrow WW \rightarrow  e \met 2j$, $h
\rightarrow ZZ \rightarrow ee2j$, $Wh \rightarrow e \met 2j$, and
$Zh \rightarrow ee2j$. See Ref.~\cite{QUAERO}.

\begin{table}[htb]
\centering
\begin{tabular}{|c|c|}
\hline\hline
 {\bf Final state}  & {\bf Considered variable}\\
\hline\hline $\met$ & $\met$ \\
\hline
$\ge$ $1$ Charged Leptons &
$\sum{p_T^\ell}$\\
\hline
$\ge$ $1$ Vector Bosons & $\sum{p_T^{\gamma/W/Z}}$ \\
$\ge$ $1$ jets & $\sum'{p_T^j}$ \\\hline\hline
\end{tabular}
\caption{A quasi-model-independently motivated list of interesting
variables for any final state.  The set of variables to consider
for any particular final state is the union of the variables in
the second column for each row that pertains to that final state.}
\label{tbl:variables}
\end{table}

\section{Conclusions}
We presented a sample of the latest results on physics beyond the SM at
Tevatron. At present both CDF and D$\not$O are collecting data with
upgraded detectors challenging the Tevatron performances.
With increase in luminosity and
with present improved detector performances we expect these
results to be greatly extended.

\section{Acknowledgments}
I want to thank the Organizers of
{\it Incontri sulla Fisica delle Alte Energie} for the
excellent conference and for their kind hospitality.


\eject
\end{document}